\documentclass[epjCONF]{svjour}
\usepackage{graphicx}
\usepackage[varg]{txfonts} % Times fonts
\usepackage[latin1]{inputenc}
\session-title{%
19$^{\textnormal{\footnotesize th}}$ International %
IUPAP Conference on Few-Body Problems in Physics%
}
\begin{document}
\title{%
%Insert your title here: 
Nonperturbative Renormalisation Group: Applications to the few and many-body systems
}%
\author{%
% add email of the responsible author:
B.Krippa \inst{1,2}\fnmsep\thanks{\email{boris.krippa@manchester.ac.uk}} 
%N.N. Surname2\inst{2} 
%\and %
%N.N. Surname3\inst{1} 
}
\institute{%
School of Physics and Astronomy, University of Manchester, Manchester M13 9PL, UK
\and %
Institute for Theoretical and Experimental Physics, 117259, Moscow, Russia
}
\abstract{
We consider the applications of functional renormalisation group to few and many-body systems.
As an application to the few-body dynamics we study the ratio between the fermion-fermion scattering length and the dimer-dimer scattering length for systems of few nonrelativistic fermions.
 We find a strong dependence on the cutoff function used
in the renormalisation flow for a two-body truncation of the action. Adding a simple three-body
term substantially reduces this dependence. In the context of many-body physics we  study the   dynamics of both symmetric and asymmetric
many-fermion systems using the same functional renormalisation technique. It is demonstrated that 
functional renormalisation group gives sensible and reliable results and provides a solid theoretical ground for the future studies.  Open questions as well as lines of further developments are discussed.
}
\maketitle
%
% 
%----- Beginning of MAIN TEXT  --------------------------------------- 
% 
Nonperturbative/Exact Renormalisation
Group (ERG) has become a popular tool to study strongly interacting dynamics \cite{Wet}. 
It has been used to study  a variety of physical systems, from nonrelativistic two-body
\cite{Bir} and many-body \cite{Kri} systems to effective quark models
\cite{Wet1} and gauge theories \cite{Gie}. The main object of the
ERG is the scale-dependent average effective action $\Gamma_{k}$,
where $k$ is an auxiliary running scale (see the reviews \cite{Wet2}
for detailed discussion). The evolution of $\Gamma_{k}$ with the
running scale is described by a flow equation. The average effective
action at scale $k^{2}$ contains the fluctuations with momenta $q^{2}$
larger than $k^{2}$. Fluctuations with momenta smaller than $k^{2}$
are integrated out. In the limit $k\rightarrow0$ all the fluctuations
are included and the full effective action is recovered. To integrate
out modes with $q^{2}<k^{2}$ one introduces a cutoff function $R_{k}(q)$
in path integral. This function should vanish in the limit $k\rightarrow0$
to ensure the physical results to be cutoff independent and behave
like $k^{\alpha}$, $\alpha>0$ when $q\rightarrow0$. In the limit
of large $k$ the average effective action should turn into the basic
action.
 
In this paper we consider several applications of ERG
to the physical systems. Namely, we calculate  the scattering length of two dimers -  bound states of two fermions,
which play an  important role in dynamics of ultra-cold and dilute Fermi gases.
We also consider   both symmetric and asymmetric many-fermion systems with pairing interaction and analyse
some  physical characteristic of these systems such as pairing gap, chemical potential and ground state energy.

The field of  ultra-cold  Fermi gases has become one of the hot topics in atomic physics.
 An important feature of such systems is superfluidity
which is the result of attractive fermion-fermion interactions leading
to pairing. Recent advances in using Feshbach resonances allow a tuning
of the fermion-fermion scattering length $a$. For negative scattering
length we get the small-coupling BCS state. For positive values of
$a$ bound states of two fermions (dimers) form and these
can be arranged into a Bose - Einstein condensate (BEC) \cite{Zwe}.
The size of dimers is determined by the fermion-fermion scattering
length and their binding energy is of order $1/a^{2}$.

The exact relation between dimer-dimer and fermion-fermion
scattering lengths $a_{B}=0.6a$ was established in \cite{Shl}
by solving the Schr\"odinger equation for two composite bosons interacting
with an attractive zero-range potential. This method, however, is
difficult to extend to the many-body case. Therefore, it is useful
to study the ratio $a_{B}/a$ in an approach which can be used
both for few and many-body problems.

 For an exact solution of the functional
RG equation we expect cut-off independent results for $a_{B}$. In practice, however some 
dependence of the results on the cut-off is inevitable. We can use this
dependence  as a measure of the completeness of our
parametrisation. We will see that the ansatz used for the effective action with two-body 
interaction is too crude and needs to be modified by including the term with the 
three-body interaction.

The flow of the scale-dependent effective
action satisfies 

\begin{equation}
\partial_{k}\Gamma=-\frac{i}{2}\,STr\left[(\partial_{k}R)\,(\Gamma^{(2)}-R)^{-1}\right].\end{equation}
 where $\Gamma^{(2)}$ is the second functional derivative taken with
respect to the fields entering the action, and $R$ is an operator
that drives the RG evolution. The operation $STr$ denotes the supertrace
\cite{Ma} taken over both energy-momentum variables and internal
indices and is defined as 
\begin{equation}
STr\left(\begin{array}{cc}
A_{BB} & A_{BF}\\
A_{FB} & A_{FF}\end{array}\right)=Tr(A_{BB}) - Tr(A_{FF}).
\end{equation}
 In the case of mixed boson-fermion systems $\partial_{k}R(q)$ acts
on both boson or fermion degrees of freedom. As mentioned above the flow equation is
a functional differential equation, and since there are no general
methods to solve such equations numerically, we must resort to approximations.
One approach is to parametrise the effective action by a finite set
of coefficients, turning the evolution into a system of coupled ordinary
differential equations which can then be solved numerically. In this
paper, where we study possible truncations for the system of a few
interacting s-wave fermions, our choice of ansatz for effective action
is motivated by both studies of many-body systems \cite{Kri} and
few-fermion systems in the effective field theory (EFT) approach \cite{We}.
The study of such few-body physics in the framework of ERG has just
begun. The famous Skornyakov and Ter-Martirosyan equation has been
rederived in Ref. \cite{Die1} and Efimov type of physics has been
considered in Ref. \cite{Mor} (see also \cite{Mor1}). Here we follow Ref.\cite{KWB}

We first consider the case when fermions interact pairwise. This case
was previously considered  by the Heidelberg group \cite{Die2} and more recently in \cite{Bir},
 where many the technical details can be found. Here we
just give a brief account of the formalism, concentrating on the dimer-dimer
scattering length. We reproduce the results from Ref. \cite{Die2} which allow us to establish the framework
for examining the cutoff scale dependence of the results. At some starting scale we demand the microscopic
action to be a purely fermionic theory with the contact 4-fermion
interaction without derivatives. This kind of interaction has been
extensively used in the EFT-based studies of nucleon-nucleon forces
\cite{We}. It is convenient to modify the theory by introducing an
auxiliary composite boson field. Using the Hubbard-Stratonovich transformation
the 4-fermion interaction gets replaced by a {}``Yukawa'' type of
coupling between the fermions and an auxiliary boson. A kinetic term
for the boson is then generated in the RG evolution. We also add a
local three-body interaction, which we expect to be generated during
the evolution, to the standard two-body result. This can be captured
in a following parametrisation of the effective action,
 
\begin{eqnarray}
\Gamma[\psi,\psi^{\dagger},\phi,\phi^{\dagger},k]=\Gamma_{2b}[\psi,\psi^{\dagger},\phi,\phi^{\dagger},k] \nonumber \\
  -\lambda\int d^{4}x\,\psi^{\dagger}(x)\phi^{\dagger}(x)\phi(x)\psi(x),
\end{eqnarray}
where
\begin{eqnarray}
\Gamma_{2b}[\psi,\psi^{\dagger},\phi,\phi^{\dagger},k] =\int d^{4}x\,\Biggl[\int d^{4}x'\,\phi^{\dagger}(x)\Pi(x,x';k)\phi(x')\nonumber \\
 +\psi^{\dagger}(x)\left(i\partial_{t}+\frac{1}{2M}\,\nabla^{2}\right)\psi(x)\nonumber \\
 -g\biggl(\frac{i}{2}\,\psi^{{\rm T}}(x)\sigma_{2}\psi(x)\phi^{\dagger}(x)\label{eq:ansatz}
 -\frac{i}{2}\,\psi^{\dagger}(x)\sigma_{2}\psi^{\dagger{\rm T}}(x)\phi(x)\biggr)\Biggr].
\end{eqnarray}

We shall first concentrate on the two-body part $\Gamma_{2b}$.
Here $\Pi(x,x';k)$ is the scale-dependent boson self-energy. The
evolution of this self-energy is given by \begin{equation}
\partial_{k}\Pi(x,x';k)=\frac{\delta^{2}}{\delta\phi(x')\delta\phi^{\dagger}(x)}\partial_{k}\Gamma|_{\phi=0},\end{equation}
but from here-on we shall express all evolution in momentum space.
Note that only fermion loops, which only depend on the fermion cut-off
$R_{F}$ and its derivatives, contribute to the evolution of the boson
self-energy in vacuum. Integrating the ERG equation with the fermionic
cut-off 
\begin{equation}
R_{F}({\vec{q}},k)=\frac{k^{2}-q^{2}}{2M}\theta(k-q),\label{eq:fermcut-off}
\end{equation}
and the renormalisation condition that the constant term in $\Pi(P_{0},P;0)$
reproduces the inverse of the zero energy $T$ matrix, we find 
\begin{equation}
\Pi(P_{0},P;K)=\frac{g^{2}M}{4\pi^{2}}\left[-\frac{4}{3}K+\frac{\pi}{a}+\frac{16}{3K}\left(MP_{0}-\frac{P^{2}}{2}\right)-\frac{P^{3}}{24K^{2}}+...\right].\end{equation}
 The on-shell fermion-fermion scattering amplitude in the physical
limit $k\rightarrow0$ is given by \begin{equation}
\frac{1}{T(p)}=\frac{1}{g^{2}}\Pi(P_{0},P;0),
\end{equation}
 where $p=\sqrt{2MP_{0}-P^{2}/2}$ is the relative momentum of two
fermions and $P_{0}(P)$ denote the total energy (momentum) flowing
through the system. From the gradient expansion, we define boson wave-function
and mass renormalisation factors by

\begin{equation}
Z_{\phi}(k)=\frac{\partial}{\partial P_{0}}\Pi(P_{0},\vec{P};k)\Biggr|_{P_{0}=\mathcal{E}_{D},\vec{P}=0},
\end{equation}
 and

\begin{equation}
\frac{1}{4M}Z_{m}(k)=-\frac{\partial}{\partial P^{2}}\Pi(P_{0},\vec{P};k)\Biggr|_{P_{0}=\mathcal{E}_{D},\vec{P}=0}.\end{equation}
Here $\mathcal{E}_{D}=-1/(Ma^{2})$ which is the bound-state energy
of a pair of fermions. Note that these renormalisation factors are
only identical in vacuum for a limited subset of cutoff functions   preserving Galilean invariance to
 otherwise the identity $Z_{\phi}(k)=Z_{m}(k)$ holds
only in the physical limit $k\rightarrow0$ and the evolution of these
renormalisation factors should be calculated separately. 

The evolution of the boson-boson scattering amplitude follows from

\begin{equation}
-\frac{2}{(2\pi)^{4}}\partial_{k}u_{2}(\mathcal{E}_{D},k)=\frac{\delta^{4}}{\delta\phi^{2}(\mathcal{E}_{D},0)\delta\phi^{\dagger2}(\mathcal{E}_{D},0)}\partial_{k}\Gamma|_{\phi=0}.\end{equation}
 The evolution can be separated into fermionic and bosonic contributions
containing $\partial_{k}R_{F}$ and $\partial_{k}R_{B}$, respectively.
We first look at the mean-field result, where boson contributions
are neglected. The evolution of $u_{2}$ is then given by \begin{equation}
\partial_{k}u_{2}=-\frac{3g^{4}}{4}\int\frac{d^{3}{\vec{q}}}{(2\pi)^{3}}\,\frac{\partial_{k}R_{F}}{\left[(E_{FR}(\vec{q},k)-\mathcal{E}_{D}/2\right)]^{4}},\end{equation}
 where $E_{FR}(\vec{q},k)=\frac{1}{2M}\, q^{2}+R_{F}(q,k)$. Explicit
calculations give \begin{equation}
u_{2}(0)=\frac{1}{16\pi}M^{3}g^{4}a^{3}\end{equation}
 where we have again used the sharp cut-off function of the form Eq.~(\ref{eq:fermcut-off}).

The mean field scattering amplitude at threshold is \begin{equation}
T_{BB}=\frac{8\pi}{2M}a_{B}=\frac{2u_{2}(0)}{Z_{\phi}^{2}}=\frac{8\pi a}{M}.\end{equation}
 This is the well-known mean-field result $a_{B}=2a$ \cite{Ha}
which is far from the exact value $a_{B}=0.6a$ \cite{Shl}. This
implies that beyond-mean-field effects such as dimer-dimer rescattering
are important to be considered. To include these effects one needs to
take into account the boson loops. After some algebra we get

\begin{equation}
\partial_{k}u_{2} = \partial_{k}u_{2}|_{F} + \partial_{k}u_{2}|_{B}
\end{equation}
Here
\begin{equation}
\partial_{k}u_{2}|_{B}=\frac{u_{2}^{2}(k)}{2Z_{\phi}^{3}(k)}\int\frac{d^{3}{\vec{q}}}{(2\pi)^{3}}\,\frac{\partial_{k}R_{B}}{\left[E_{BR}(\vec{q},k)-\mathcal{E}_{D}\right]{}^{4}}
\end{equation}
and
\begin{equation}
\partial_{k}u_{2}|_{F} = -\,\frac{3 g^4}{4}\int\frac{d^3{\vec q}}{(2\pi)^3}\,
\frac{\partial_kR_F}{\left[E_{FR}(\vec{q},k)-\mathcal{E}_{D}/2\right]{}^{4}} 
\end{equation}
 where 
\begin{equation}
E_{BR}(\vec{q},k)=\frac{1}{4M}\, q^{2}+\frac{u_{1}(k)}{Z_{\phi}(k)}+\frac{R_{B}(q,k)}{Z_{\phi}(k)}
\end{equation}
 and 
\begin{equation}
u_{1}(k)=-\Pi(\mathcal{E}_{D},0;k).
\end{equation}
 The boson cut-off function is to be chosen as close as possible to
the fermionic one, \begin{equation}
R_{B}(\vec{q},k)=Z_{\phi}\frac{(c_{B}k)^{2}-q^{2}}{2M}\theta(c_{B}k-q),\end{equation}
apart from the addition of a parameter $c_{B}$, which sets the relative
scale of the fermion and boson regulators, and a factor of $Z_{\phi}$.
The latter has two main advantages: Firstly, it leads to universality,
where all contributions to a single evolution equation decay with
the same power of $k$ for large $k$ and secondly we get $a_{F}$
scaling, where all terms in a single evolution equation have the same
dependence on $a_{F}$.

\begin{figure}
\begin{centering}
\includegraphics[clip,width=7cm]{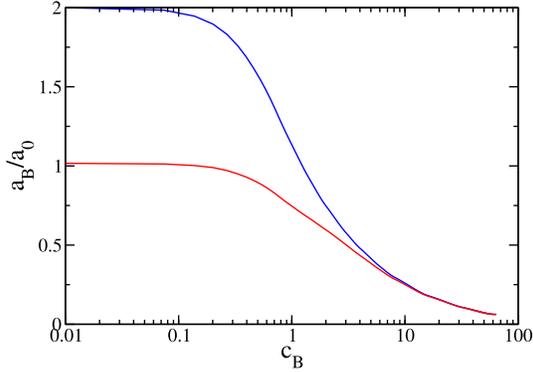} 
\par\end{centering}

\caption{Ratio of boson-boson to fermion-fermion scattering length as a function
of the scale parameter $c_{B}$. The blue (solid) curve shows results
without 3 body forces, the red (dashed) curve contains the effect
of a 3-body force. \label{fig:Ratio}}

\end{figure}

The mean-field result is recovered when $c_{B}=0$. The limit of $c_{B}\rightarrow\infty$
leads to $a_{B}\rightarrow0$. Using $c_{B}=1$ gives the value of
the ratio $a_{B}/a=1.13$. Choosing $c_{B}=\sqrt{2}$ results in $a_{B}/a_{F}=0.75$
as in Ref.~\cite{Die2}. In general, the results show rather strong
dependence on the boson scale parameter $c_{B}$ (see Fig. \ref{fig:Ratio}).
This is unwanted as it means that the physical results depend strongly
on the choice of regulator. As announced above we shall now consider the effect
of the three-body force (\ref{eq:ansatz3}).

The equations for $u_{1}$ and $Z_{\phi}$ remain unchanged but the
one for $u_{2}$ gets modified by the term describing the 3-body interaction :

\begin{equation}
\partial_{k}u^{3B}_{2}= -2\lambda g^{2}\int\frac{d^{3}{\vec{q}}}{(2\pi)^{3}}\,\frac{\partial_{k}R_{F}}{\left[E_{FR}(\vec{q},k)-\mathcal{E}_{D}/2\right]^{3}}
\end{equation}
 where we denote $E_{FR,P_{B}}=E_{FR}-P_{B}/2$ and
$E_{BR,P_{B}}=E_{BR}-P_{B}$. The evolution equation for $\lambda$
is defined by an expansion at the energy of the bound state pole for
bosons, and half of that for fermions, 

\begin{equation}
\partial_{k}\lambda  =-\frac{i}{2}\frac{\delta^{4}STr\left[\partial_{k}R(\Gamma^{(2)}-R)^{-1}\right]}{\delta\phi^{\dagger}(\mathcal{E}_{D},\vec{0})\delta\phi(\mathcal{E}_{D},\vec{0})\delta\psi^{\dagger}(\mathcal{E}_{D}/2,\vec{0})\delta\psi(P_{B}/2,\vec{0})}
\end{equation}

There are three distinct contributions to the running of $\lambda$
coming from a ladder, triangle and box diagrams.
We denote the corresponding driving terms as $D_{b}$, $D_{l}$ and
$D_{t}$, splitting the last two terms in fermionic and bosonic contributions.
After evaluation of traces and contour integrals we get 
\begin{equation}
D_{l}  =\lambda^{2}\int\frac{d^{3}{\vec{q}}}{(2\pi)^{3}}\,\frac{\partial_{k}R_{F}Z_{\phi}+\partial_{k}R_{B}}{(E_{FR,P_{B}}Z_{\phi}+E_{BR,P_{B}})^{2}},
\end{equation}
\begin{equation}
D_{t}^{F}  = g^{2}\lambda\int\frac{d^{3}{\vec{q}}}{(2\pi)^{3}}\,\frac{\partial_{k}R_{F}(E_{BR,P_{B}}+2Z_{\phi}E_{FR,P_{B}})}{E_{FR,P_{B}}^{2}(E_{FR,P_{B}}Z_{\phi}+E_{BR,P_{B}})^{2}}
\end{equation}
\begin{equation}
D_{t}^{B}  = g^{2}\lambda\int\frac{d^{3}{\vec{q}}}{(2\pi)^{3}}\,\frac{\partial_{k}R_{B}}{E_{FR,P_{B}}(E_{FR,P_{B}}Z_{\phi}+E_{BR,P_{B}})^{2}},
\end{equation}
\begin{equation}
D_{b}^{F}  =\frac{g^{4}}{4}\int\frac{d^{3}{\vec{q}}}{(2\pi)^{3}}\,\frac{\partial_{k}R_{F}(2E_{BR,P_{B}}+3Z_{\phi}E_{FR,P_{B}})}{E_{FR,P_{B}}^{3}(E_{FR,P_{B}}Z_{\phi}+E_{BR,P_{B}})^{2}}
\end{equation}
\begin{equation}
D_{b}^{B}  = \frac{g^{4}}{4}\int\frac{d^{3}{\vec{q}}}{(2\pi)^{3}}\,\frac{\partial_{k}R_{B}}{E_{FR,P_{B}}^{2}(E_{FR,P_{B}}Z_{\phi}+E_{BR,P_{B}})^{2}}.
\end{equation}
 We denote $E_{FR,P_{B}}=E_{FR}-P_{B}/2$ and
$E_{BR,P_{B}}=E_{BR}-P_{B}$ with $P_{B}$ being some external energy.
 The evolution equation for $\lambda$
is defined by an expansion at the energy of the bound state pole for
bosons, and half of that for fermions, 
Note that the the evolution of the three-fermion term $\lambda$ does
not depend on the four-fermion term interaction $u_{2}$, but $u_{2}$
depends on $\lambda$. We now need an initial condition for $\lambda$.
At infinite $k$, there is no fundamental three-fermion interaction,
and thus $\lambda=0$. To simplify the calculations we have assumed
that $\lambda=0$ is also zero at the starting scale, since $\lambda$
behaves like $1/k^{2}$ this can be used reliably.

Now we turn to the results. We first note that the results are numerically
independent of the choice of starting scale provided it is chosen
to be at least $ka_{F}\simeq100$. For $k\gg1/a$ the system is
in the ``scaling regime'' and evolves near the fixed point until
the scale becomes comparable with $1/a$. This can easily be seen
if we switch to dimensionless coupling $k^{2}\lambda$ and analyse
its evolution.

We have found that the ratio of $a_{B}/a$ decreases when the
three-body term is included. For example, assuming $c_{B}=1$ leads
to the value $a_{B}/a=0.74$ compared with $a_{B}/a=1.13$
without inclusion of the the three-body term, and the choice $c_{B}=\sqrt{2}$
as used in \cite{Die2} gives $a_{B}/a=0.69$. We have shown in Fig.1
the behaviour of $a_{B}/a$ as a function of the scale $c_{B}$.

The upper curve corresponds to the calculations without three-body
term and the lower curve gives the results of the full calculations
with three-body term included. It is seen from Fig.1 that taking into
account the three-body forces results in significantly weaker dependence
of $a_{B}/a$ on the scale $c_{B}$. Note that at large $c_{B}$,
which corresponds to integrating out fermions first, the two curves
approach each other. Indeed, in this limit the dominant contribution
to $a_{B}/a$ comes from the bosonic part (proportional to $R_{B}$)
of the equation for $u_{2}$ which does not depend on the three-body
coupling $\lambda$. On the other hand, with small $c_{B}$ the main
contribution comes from the fermion loops which get modified when
the three-body coupling $\lambda$ is included. We expect this picture
to be qualitatively correct for any type of boson regulator although
the quantitative details will vary depending on the functional form
of the regulator used.

In spite of showing a clear improvement over the calculations with
the two-body interactions only, adding the simplest possible three-body
term is not enough to ensure the complete (or almost complete) independence
of the results from the scale parameter $c_{B}$ in the region $c_{B}\simeq1$.
It is worth emphasising again that, as long as any truncation of the
effective action is made, the results will never be completely independent
of the scale $c_{B}$. Instead, we expect some {}``stability window''
where the slope of $a_{B}/a$ is small. It is somewhat similar
to the QCD sum rules \cite{WZS}, where the physical observables are
expected to be independent of an auxiliary parameter called Borel
mass. It seems likely that further extensions of the effective action
will result in stability of the results with respect to the variations
of $c_{B}$ in wider region. Such extensions may include both four-body
interactions as well as energy and/or momentum dependent three-body
forces. Work along these lines is now in progress.

We next consider the applications of the ERG to the  finite density many-fermion systems.
It is well established that in the medium  even a weak attraction between fermions
 may lead to the intrinsically nonperturbative  phenomenon,  the superfluidity,  characterised  by
  rearrangement of the ground
state and appearance of the gap in the spectrum.
 The fermions form correlated pairs which, depending on the strength of the interaction, may lead  to different physical regimes.
The weak coupling regime (BCS phase) corresponds to a pair with the spatial size much larger then the radius of the interaction 
so that no actual
bound two-body subsystem is formed, while in the strong regime corresponding to the Bose-Einstein Condensation (BEC) the fermion pairs form 
compact deeply bound two-body states. We demand that at high scale our theory  be a purely fermionic theory with the contact interaction described by  the lagrangian 
\begin{equation}
{\cal L}_i=-\frac{1}{4}\,C_0\left(\psi^\dagger\sigma_2\psi^{\dagger{\rm T}}\right)
\left(\psi^{\rm T}\sigma_2\psi\right),
\end{equation}
which is identical to one, used to calculate the fermion-fermion scattering in vacuum so we can link  the vacuum and in-medium ERG calculations.
This is what makes the approach universal. Since we are interested in the appearance of the correlated fermion pairs in a physical 
ground state, we need to parametrise our effective action in terms of corresponding variables.
A natural way to do this is to introduce a boson field whose vacuum 
expectation value (VEV) describes this correlated pair   and study the evolution of this effective degrees of freedom. At  
the start of the RG evolution, the boson 
field is not dynamical and is introduced through a 
Hubbard-Stratonovich transformation of the four-point interaction.
As we integrate out more and more of the fermion degrees of freedom by 
running $k$ to lower values the  dynamical term in the bosonic
effective action is generated.

 The corresponding ansatz 
for the  boson-fermion effective action can be written as
\begin{eqnarray}
\Gamma[\psi,\psi^\dagger,\phi,\phi^\dagger,k]&=&\int d^4x\,
\left[\phi^\dagger\left(Z_\phi\, (i \partial_t +2\mu)
+\frac{Z_m}{2m}\,\nabla^2\right)\phi-U(\phi,\phi^\dagger)\nonumber\right.\\
\noalign{\vskip 5pt}
&&\qquad\qquad+\psi^\dagger\left( Z_\psi (i \partial_t+\mu)
+\frac{Z_M}{2M}\,\nabla^2\right)\psi\nonumber\\
\noalign{\vskip 5pt}
&&\qquad\qquad\left.- g \left(\frac{i}{2}\,\psi^{\rm T}\sigma_2\psi\phi^\dagger
-\frac{i}{2}\,\psi^\dagger\sigma_2\psi^{\dagger{\rm T}}\phi\right)\right].
\label{eq:ansatz}
\end{eqnarray}
Here $M$ is the mass of the fermions in vacuum and the factor $1/2m$ in the 
boson kinetic term is chosen simply to make $Z_m$ dimensionless. The
couplings, the chemical potential $\mu$,  the 
wave-function renormalisations $Z_{\phi,\psi}$ and the kinetic-mass 
renormalisations $Z_{m,M}$ all run with  $k$, the scale of the regulator. 
The bosons are , in principle coupled to the chemical potential via a quadratic term in $\phi$, 
but this can be absorbed into the potential by defining $\bar U=U-2\mu Z_\phi\phi^\dagger\phi$.
We expand this potential about its minimum, $\phi^\dagger\phi=\rho_0$, so that the coefficients $u_i$ are defined at $\rho=\rho_0$,  
\begin{equation}
\bar U(\rho)= u_0+ u_1(\rho-\rho_0)
+\frac{1}{2}\, u_2(\rho-\rho_0)^2
+\frac{1}{6}\, u_3(\rho-\rho_0)^3+\cdots,
\label{eq:potexp}
\end{equation}
where we have introduced $\rho=\phi^\dagger\phi$. The phase of the system is determined by the coefficient $u_1$. In the symmetric phase
we have $\rho_0=0$ so that the expansion takes the form 
\begin{equation}
\bar U(\rho)= u_0+u_1\rho+\frac{1}{2}\, u_2\rho^2+\cdots.
\label{eq:potexps}
\end{equation}
The potential in the condensed phase 
can be simplified to
\begin{equation}
\bar U(\rho)= u_0+\frac{1}{2}\, u_2(\rho-\rho_0)^2+\cdots.
\label{eq:potexpc}
\end{equation}
In our current work we shall truncate this potential at quartic order in the field
(order $\rho^2$).
We treat the wave function renormalisation factor for the bosons
in the same way, expanding it about $\rho=\rho_0$ as
\begin{equation}
Z_\phi(\rho)= z_{\phi 0}+ z_{\phi 1}(\rho-\rho_0)+\cdots.
\label{eq:Zphiexp}
\end{equation}
 The other couplings and renormalisation
factors can be treated similarly.

The fermions are not dressed at this point, the bosons are just auxiliary fields and therefore we can assume that  
$Z_\psi(K)=1$ and  $Z_M(K)=1$. After rather lengthy algebra one can obtain the explicit expressions for the couplings
in the effective action. As an example we show below the corresponding ERG equation  for the effective potential.

\begin{eqnarray}
\partial_k \bar U=
&=&-\,\frac{1}{Z_\psi}\int\frac{d^3{\vec q}}{(2\pi)^3}\,\frac{E_{FR}}
{\sqrt{E_{FR}^2+\Delta^2}}\,\partial_kR_F\nonumber\\
\noalign{\vskip 5pt}
&&+\,\frac{1}{2Z_\phi}\int\frac{d^3{\vec q}}{(2\pi)^3}\,
\frac{E_{BR}}{\sqrt{E_{BR}^2-V_B^2}}
\,\partial_kR_B.\label{eq:potevol}
\end{eqnarray}
where
\begin{equation}
E_{BR}(q,k)=\frac{Z_m}{2m}\,q^2+u_1
+u_2(2\phi^\dagger\phi-\rho_0)+R_B(q,k),\quad V_B= u_2\phi^\dagger\phi.
\end{equation}
The  evolution equations for the coefficients in our 
expansion of the effective potential are obtained from the derivatives of
$\partial_k \bar U$ with respect to $\rho=\phi^\dagger\phi$.

The initial conditions for them can be fixed by demanding that at high scale the evolution of the system is close
to the vacuum one.One could introduce a cut-off function that tends to a $p_F$-independent form
for $K\gg p_F$. However in practice a modification of the renormalisation
procedure is more convenient. In the region $K\gg p_F$, we can ignore boson
loops. The evolution of
quantities such as $u_1(p_F,K)$, $u_2(p_F,K)$, $Z_\phi(p_F,K)$ and
$Z_m(p_F,K)$ is thus similar to 
the vacuum case, except for the different cut-off. This allows us to define
$u_1(p_F,K)$ to be
\begin{equation}
\frac{u_1(p_F,K)}{g^2}=-\frac{M}{4\pi a}
+\frac{1}{2}\,\int\frac{d^3{\vec q}}{(2\pi)^3}\,
\left[\frac{1}{E_{FR}(q,0,0)}-\frac{sgn(q-p_F)}{E_{FR}(q,p_F,K)}\right]\, .
\label{eq:u1Kfull}
\end{equation}
This expression can be thought of as being generated by the vacuum evolution 
using a modified cut-off that interpolates smoothly between $R_F(q,p_F,k)$
for $k\gg p_F$ and $R_F(q,0,k)$ for $k\simeq p_F$. It ensures that our 
renormalised parameter $u_1(p_F,K)$, defined using $R_F(q,p_F,k)$ for large 
$k$, corresponds to the physical scattering length in vacuum.

The initial values for $u_2(p_F,K)$, $Z_\phi(p_F,K)$ and $Z_m(p_F,K)$ 
can be determined using similar procedures, although this is not so 
crucial since these quantities do not contain linearly divergent pieces 
and so all their $p_F$-dependence is suppressed by powers of $p_F/K$.
One convenient choice is to take their starting values to be zero at some
large but finite scale $K$. An alternative is to require that they tend 
to zero as $K\rightarrow\infty$.

The initial condition for the energy density is most conveniently
expressed in terms of $\tilde u_0$ which,  in the symmetric phase, is
simply given by the energy of a free Fermi gas, measured relative to the 
chemical potential, and so its initial value 
is just
\begin{equation}
\tilde u_0(K)=2\int\frac{d^3{\vec q}}{(2\pi)^3}\,E_{FR}(q,p_F,0)\theta(p_F-q).
\end{equation}

Now we turn to the results which are shown on  Fig. 2. First we note that  the value of the
physical gap  is  practically independent of 
the starting scale $K$ provided $K > 5 fm^{-1}$.

\begin{figure}
\begin{center}
\includegraphics[width=9cm, clip]{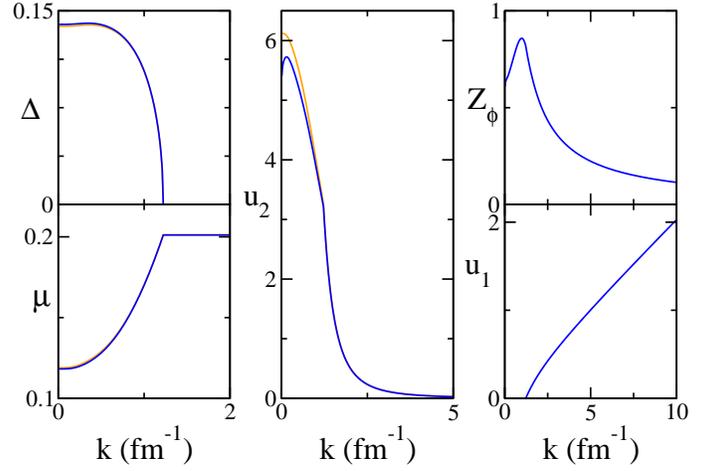}
\end{center}
\caption{\label{fig.2}Numerical solution of the evolution equations for infinite scattering length.The thinner curve (orange online) corresponds to the mean-field
calculations and the thicker one (blue online) includes boson loops. The difference however is noticeable only for the evolution of $u_2$ (graph in the middle). }
\end{figure}

 At starting scale
 the system is in the symmetric phase and remains in this phase until $u_1$ hits zero at $k_{crit}\simeq 1.2 fm^{-1}$ where the artificial second
order phase transition to a broken phase occurs and the energy gap is formed. Already at $k \simeq 0.5$ the running scale has essentially no effect on
 the gap. 
The results obtained for the gap correspond to the case of the infinite negative scattering length.
 To study the BCS-BEC crossover we have to solve the evolution equation for a wider range of the scattering lengths, including the positive values.
The corresponding results for the evolution of chemical potential as a function of the parameter $p_{F}a$ are  shown in Fig.3.

\begin{figure}
\begin{center}
\includegraphics[width=6cm, clip]{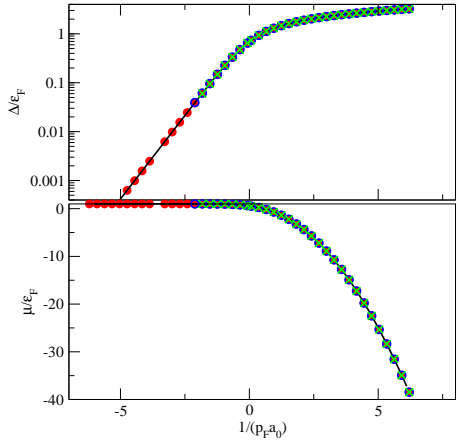}
\end{center}
\caption{\label{fig.3}Evolution of the chemical potential}
\end{figure}
While the vacuum scattering length is large and negative, the system is in the BCS-like phase with positive chemical potential, whereas if the 
 scattering length is chosen to be large and positive  reflecting the existence of a bound state near threshold
the system ends up being the collection of weakly overlapping tightly bound pairs with negative chemical potential.

As in the case of dimer-dimer scattering we expect here some dependence on the regulator used. By analogy with the vacuum case 
the inclusion of the three-body forces will probably 
make the results more stable although the corresponding calculations are much more involved compared to the vacuum case. 
Work along this line is in progress as well.

We have also calculated the ground state energy of the many-fermion system in the unitary regime - the idealised case of the infinite 
fermionic scattering length \cite{Kr1}.
In this regime the dynamics of the many-body system becomes independent of the microscopical details of the underlying interaction between two fermions.
In the $a\rightarrow - \infty$ limit the ground state energy per particle is proportional to that of the non-interacting Fermi gas.

\begin{equation}
E_{GS}=\xi E_{FG}=\xi \frac{3}{5}\frac{k^{2}_{F}}{2 M}= \xi \frac{3}{5} E_{F},
\end{equation}
where $M$ and $k_F$ are the fermion  mass and Fermi momentum correspondingly and   $\xi$ is the universal
 proportionality constant, which does not depend on the details of the interaction or type of fermions.
 The other dimensional characteristics of the cold Fermi-gas in the UR such as paring energy $\Delta$ or chemical potential $\mu$
can also be represented in the same  way

\begin{equation}
\mu = \eta E_{FG} , \qquad \Delta = \epsilon E_{FG}
\end{equation}
 The infinite scattering length implies nonperturbative treatment that why the ERG formalism is a very useful tool here. 
We found the values of $0.62, 0.96$ and $1.11$ for the universal coefficients $\xi, \eta$
and $\epsilon$ correspondingly.

The obtained value for  $\xi$ is close  to
 the experimental data  from \cite{Ge},  $\xi = 0.74(7)$.
The other measurements give $\xi = 0.34(15)$ and $\eta = 0.6(15)$\cite{Bo}; $\xi = 0.32^{+0.13}_{-0.10}$ and $\eta = 0.53^{+0.13}_{-0.10}$ \cite{Ba}; 
$\xi = 0.46(5)$   and $\eta = 0.77(5)$\cite{Pa};
$\xi = 0.51(4)$ and $\eta = 0.85(4)$ \cite{Ki}. The results of theoretical 
calculations are summarised in Table I.

\begin{table}[ht]
\caption{Universal coefficients} % title of Table
\centering % used for centering table
\begin{tabular}{c c c c} % centered columns (4 columns)
\hline\hline %inserts double horizontal lines
Ref & $\xi$ & $\epsilon$ & $\eta$ \\ [0.5ex] % inserts table
%heading
\hline % inserts single horizontal line
\cite{Ca} & 0.42 & 0.9 & 0.71 \\ % inserting body of the table
\cite{Lee} & 0.22 &  &  \\
\cite{As} & 0.41 &  & 0.7 \\
\cite{Bu2} & 0.44 &0.93 &  \\
\cite{Ch} &  & 1.03 &  \\
\cite{So} & 0.39 &  &  \\
\cite{Se} & 0.3 & 0.66 &  \\
\cite{Di3} & 0.55 &  &  \\ [1ex] % [1ex] adds vertical space
\hline %inserts single line
\end{tabular}
\label{table:nonlin} % is used to refer this table in the text
\end{table}

 As one can see 
both experiment and numerical simulations 
do not provide the coherent value of the $\xi$ constant so it is difficult to judge the quality of the numerical estimates 
provided by the  ERG calculations. One may only conclude that the ERG approach leads to the sensible values of the  universal coefficients
consistent with the experiment and lattice calculation but more detailed comparison  can be done when more accurate date are obtained.
We note, however that the value of  the universal parameters  are still somewhat higher then the ``world average''. One possible cause could 
be the neglection of the screening effects \cite{Go} which are known to decrease the values of the gap and energy density. 
Naive extrapolation of our results using the known value of the Gorkov - Melik-Barkhudarov's correction \cite{Go} indeed brings the values of the universal 
coefficients closer to the ``world average'' of the lattice and experimental data. Clearly,  this point requires  further 
analysis.

We now turn to the case of asymmetric many-fermion system. This asymmetry can be provided by unequal masses,
 different densities and/or chemical potentials. Understanding   the pairing mechanism in such settings  would  be of immense value for different many fermion systems
from atomic physics to strongly interacting quark matter. The important theoretical issue  to be resolved here  is the nature of the ground state. Several 
competing states have been proposed so far. These include: LOFF \cite{Lar} phase, breached-pair (BP)
superfluidity \cite{Wil} (or Sarma phase) and mixed phase \cite{Bed}. Establishing  the true ground state  is still an open question.  
It was shown, for example,   that LOFF and mixed  phases are more stable then the Sarma phase in the systems of fermions with the mismatched Fermi surfaces 
and with both equal and different masses \cite{Cal}. We concentrate on the case of two fermion species.
 The ansatz for the effective action we use to run the ERG evolution is a simple generalisation of 
the one used for the many-fermion system with one type of fermions and consists of summing over two types of fermions. Calculating the second functional derivatives, taking the matrix trace 
 and carrying out the pole integration in the loop integrals
 we get the evolution equation for $\bar U$ at constant chemical potentials
\begin{eqnarray}
\partial_k \bar U
&=&-\frac{1}{2}\,\int\frac{d^3{\vec q}}{(2\pi)^3}\,\frac{E_{F,S}}
{\sqrt{E_{F,S}^2+\Delta^2}}\,[\partial_kR_{F,a} + \partial_kR_{F,b}]\nonumber\\
\noalign{\vskip 5pt}
&&+\,\frac{1}{2Z_\phi}\int\frac{d^3{\vec q}}{(2\pi)^3}\,
\frac{E_{B}}{\sqrt{E_{B}^2-V_B^2}}
\,\partial_kR_B.\label{eq:potevol}
\end{eqnarray}

It is worth mentioning that poles 
 in the fermion  propagator occur at
\begin{equation}
q_{0}^{1,2}= - E_A \pm\sqrt{E_{S}(q,k)^2+\Delta^2}.
\end{equation}
Here 
\begin{equation}
E_S = (E_{FR,a} +E_{FR,b})/2,\qquad E_A = (E_{FR,a} - E_{FR,b})/2,
\end{equation}
 and 
\begin{equation}
E_{FR,i}(q,p_{\mu,i},k)=\frac{1}{2 M_i}\,q^2-\mu_i+R_F(q,k),  \qquad \Delta^2=g^2\phi^\dagger\phi.
\end{equation} 
and we have introduced
$p_{\mu,i}=\sqrt{2M_{i}\mu_i}$, the Fermi momentum corresponding to the (running) value of $\mu_i$.
At $k=0$ ($R_F=0$) in the condensed phase, these become
 exactly the dispersion relations obtained in \cite{Wil} where the possibility of having the gapless excitations
has been discussed. The ordinary BCS spectrum can easily be recovered when the asymmetry of the system is vanishing
($E_{A}\rightarrow 0$). 
The first term in the evolution equation for the effective potential describes the evolution of the system related to the  fermionic degrees of
 freedom whereas the second one takes into
 account the bosonic contribution. The mean field results can be recovered if the second term
 is omitted. In this case the equation
for the effective potential can be integrated analytically and then, 
by differentiating the effective potential with respect to $\rho$ and setting the derivative equal to zero, one can derive the standard mean-field gap equation.

In this paper we consider the simplified  case of two 
fermion species with the different masses and the  same Fermi momenta \cite{Kr2}.
It implies that the chemical potentials are different. In this situation neither the Sarma phase nor the LOFF phase exists and the system experiences the BCS pairing depending
however on the mass asymmetry. The general case of the mismatched Fermi surfaces will be discussed in the subsequent publication.
For simplicity we consider the case of the hypothetical ``nuclear'' matter with short range attractive interaction 
between  two types of fermions, light and heavy,
and study the behaviour of the energy gap as the function of the mass asymmetry. We choose the Fermi momentum to be  $p_{F}=1.37 fm^{-1}$.
 One notes that the formalism is applicable to any type of a many-body system with two fermion species
 from quark matter to fermionic atoms so that the    hypothetical asymmetrical ``nuclear'' matter is simply chosen as  a study case.
We assume that $M_a < M_b$, where $M_a$ is always the mass of the physical nucleon.

First we consider the case of the unitary limit where the scattering length $a = -\infty$. 
The results of our calculations for the gap 
 are shown on Fig. 4. 

\begin{figure}
\begin{center}
\includegraphics[width=8cm,  keepaspectratio,clip]{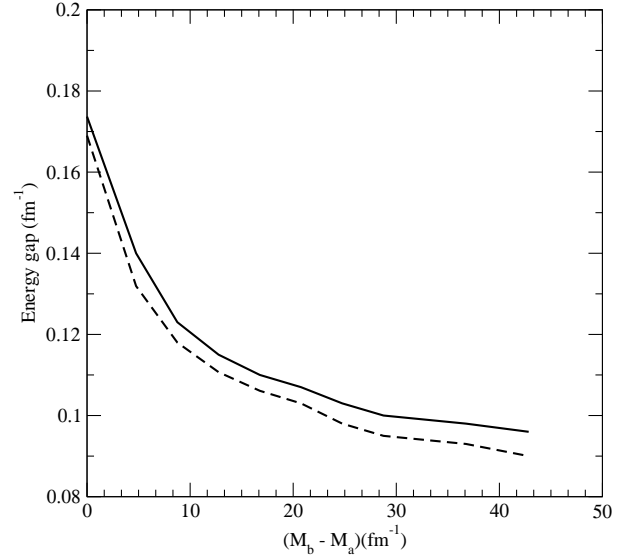}
\end{center}
\caption{\label{Fig.4}Evolution of the gap in the MF approach (dashed curve) and with boson loops (solid curve) in the unitary regime
$a = -\infty$ as a function of a mass asymmetry.}
\end{figure}

 We see from this figure  that  increasing  mass asymmetry leads to a decreasing gap that seems to be a natural result.
However, the effect of the boson loops is found to be   small. We found essentially no effect in symmetric phase, $2-4 \%$ corrections for the value of the 
gap in the broken phase and even smaller corrections for the chemical potential so that one can conclude that
the MF approach indeed provides the reliable description in the unitary limit for both small and large mass asymmetries.
It is worth mentioning that the
 boson contributions are more important for the evolution of $u_2$ where they drive  $u_2$ to zero as $k\rightarrow 0$ making the effective potential
convex in agreement with the general expectations. This tendency  retains in the unitary regime regardless of the mass asymmetry. 

We have also considered
the behaviour of the gap as the function of the parameter $p_F a$ for the cases of the zero asymmetry $M_a = M_b$ and the  maximal asymmetry $M_b = 10 M_a$.
The results are shown on Fig.5.
\begin{figure}
\begin{center}
\includegraphics[width=8cm,  keepaspectratio,clip]{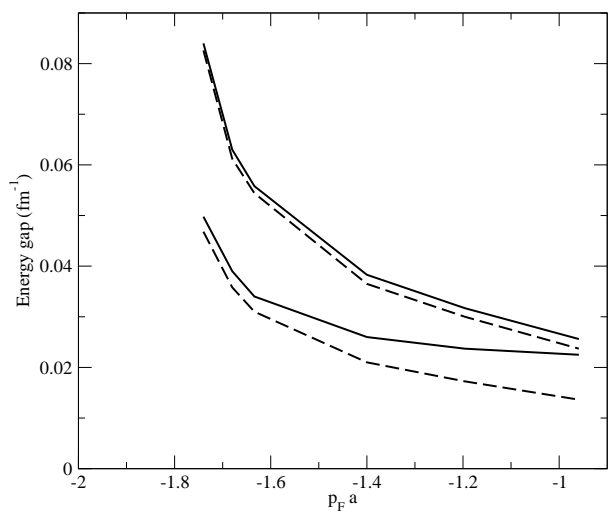}
\end{center}
\caption{\label{Fig.5}Evolution of the gap as a function of the parameter $p_F a$. The upper 
pair of the  curves corresponds to the calculations with no asymmetry in the MF approach (dashed curve) and with boson loops (solid curve) and the lower pair 
of the curves describes the results of calculations with the  maximal asymmetry when $M_b = 10 M_a$}.
\end{figure}

  One can see from  Fig.5 that in the  case of zero (or small) asymmetry  the corrections stemming from  boson loops
are small at all values of   the parameter $p_F a$ considered here (down to $p_F a = 0.94)$. On the contrary, when  $M_b = 10 M_a$ these corrections, being 
rather small at $p_F a \geq  2$ becomes significant ($\sim  40\%$) when the value of $p_F a$ decreases down to $p_F a \sim  1$. We found that  
at $p_F a \sim  1$ the effect of boson fluctuations becomes  $\sim  10\%$  already for $M_b = 5 M_a$. One can therefore conclude 
 that the regime of large mass asymmetries, which starts  approximately at  $M_b > 5 M_a$, 
 with moderate scattering length and/or the Fermi momenta is the one where the MF description becomes less accurate so that the calculations going beyond the MFA
are needed. One might expect that the deviation from the mean field results could even be stronger in a general case of a large mass asymmetry and 
the mismatched Fermi surfaces but the detailed conclusion can only be drawn after the actual calculations are performed.

In summary, we have performed the ERG analysis of a number of fermionic systems. In particular we have considered boson-boson
scattering length, where bosons are just bound states of two fermions. Our study has revealed that, starting from rather 
simple ansatz with two-body interaction term one can get the results,
close to the exact value. However, the results turn out quite sensitive
to the value of $c_{B}$, the boson scale parameter. We have shown
that the inclusion of the three-body interaction brings the ratio
$a_{B}/a$ closer to the exact value and significantly reduce
the sensitivity of the results to the boson scale parameter. We have also studied 
the many-fermion systems with attraction leading to pairing phenomena and calculated a number 
the many-body parameters such as gap and its dependence on mass asymmetry, chemical potential both in BCS and BEC regions and the ground state energy. 
The directions for further improvement of the approach are outlined.

\section{Acknowledgement}

 The author is supported by the EU FP7 programme (Grant
219533).


\begin{thebibliography}{16}

\bibitem{Wet}C. Wetterich, Phys. Lett. B\textbf{301}, 90 (1993),
T. Morris, Phys. Lett. B\textbf{334}, 355 (1994) {[}arXiv:hep-ph/9403340{]}. 


\bibitem{Bir}M. C. Birse, Phys. Rev. C\textbf{77}, 047001 (2008)
{[}arXiv:0801.2317{]}. 

\bibitem{Kri}M. C. Birse, B. Krippa, J. A. McGovern and N. R. Walet,
Phys. Lett. B\textbf{605}, 287 (2005) {[}arXiv:hep-ph/0406249{]}. 

\bibitem{Wet1}D.-U. Jungnickel and C. Wetterich, Phys. Rev. D\textbf{53},
5142 (1996) {[}arXiv:hep-ph/9505267{]}. 

\bibitem{Gie}H.Gies, arXiv:hep-ph/0611146. 

\bibitem{Wet2} J. Berges, N. Tetradis and C. Wetterich, Phys. Rept.
\textbf{363}, 223 (2002) {[}arXiv:hep-ph/0005122{]}, B. Delamotte,
D. Mouhanna and M. Tissier, Phys. Rev. B\textbf{69}, 134413 (2004)
{[}arXiv:cond-mat/0309101{]}. 

\bibitem{Zwe}M. Greiner, C. A. Regal and D. S. Jin, Nature \textbf{426},
537 (2003), doi:10.1038/nature02199; S. Jochim, M. Bartenstein, A.
Altmeyer, G. Hendl, S. Riedl, C. Chin, J. Hecker Denschlag, R. Grimm,
Science \textbf{302}, 2101 (2003), doi:10.1126/science.1093280, M.
W. Zwierlein, C. A. Stan, C. H. Schunck, S. M. F. Raupach, S. Gupta,
Z. Hadzibabic, and W. Ketterle., Phys. Rev. Lett. \textbf{91}, 250401
(2003), doi:10.1103/PhysRevLett.91.250401 . 

\bibitem{Shl}D. S. Petrov, C. Salomon and G. V. Shlyapnikov, Phys.
Rev. Lett. \textbf{93}, 090404 (2004), doi:10.1103/PhysRevLett.93.090404

\bibitem{Ma}S. Martin, {}``A Supersymmetry Primer'' in Perspectives
on supersymmetry, G. L. Kane, ed. (world Scientific, Singapore, 2008)
{[}arXiv:hep-ph/9709356{]}. 

\bibitem{We}S. Weinberg, Nucl. Phys. B\textbf{363} 3, (1991); D.
B. Kaplan, M. J. Savage and M. B. Wise, Nucl. Phys. B\textbf{534},
329 (1998); P. F. Bedaque, H. W. Hammer and U. van Kolck, Nucl. Phys.
A\textbf{676}, 357 (2000). 

\bibitem{Die1}S. Diehl, H. C. Krahl, and M. Scherer, Phys. Rev. C\textbf{78},
034001 (2008) {[}arXiv:0712.2846{]} . 

\bibitem{Mor}S. Moroz, S. Floerchinger, R. Schmidt, C. Wetterich,
Phys. Rev. A\textbf{79,} 042705 (2009) {[}arXiv:0812.0528{]}. 

\bibitem{Mor1}R.Schmidt, S. Floerchinger, C. Wetterich, Phys. Rev.
A \textbf{79}, 053633 (2009) {[}arXiv:0812.1191{]}. 

\bibitem{KWB}B. Krippa, N. R. Walet and M. C. Birse, [arXiv:0911.4608]

\bibitem{Die2}S. Diehl, H.Gies, J. M. Pawlowski, and C. Wetterich, 
Phys.Rev.\textbf{A76}, 021602 (2007){[}arXiv:cond-mat/0701198{]}.

\bibitem{Ha}R. Haussmann, Z. Phys. B \textbf{91}, 291 (1993). 

\bibitem{WZS}M. A. Shifman, A. I. Vainshtein and V. I. Zakharov,
Nucl. Phys. B\textbf{147}, 385 (1979).

\bibitem{Kr1}B. Krippa, J.Phys.\textbf{A42}, 465002 (2009). 

\bibitem{Ge}M. E. Gehm et al.,  Phys. Rev. A\textbf{68} 011401(R), (2003).

\bibitem{Bo}T. Bourdel et al., Phys. Rev. Lett. \textbf{93} 050401, (2004).

\bibitem{Ba}M. Bartenstein et al., Phys. Rev. Lett. \textbf{92}, 120401,(2004).

\bibitem{Pa}G. B. Partridge et al., Science \textbf{311}, 503, (2006).
 
\bibitem{Ki}J. Kinast et al, Science \textbf{307}, 1296, (2005).

\bibitem{Ca}J. Carlson et al., Phys. Rev. Lett. \textbf{91} 050401, (2003); 

\bibitem{Lee}D. Lee,  Phys. Rev. \textbf{B73}, 115112 (2006).

\bibitem{Bu2}A. Bulgac, J. E. Drut and P. Magierski,  Phys. Rev. A\textbf{78} 023625, (2008).

\bibitem{As}G. E. Astrakharchik et al., Phys. Rev. Lett \textbf{93} 200404, (2004).

\bibitem{Ch} S. Y. Chang et al.,  Phys. Rev. A\textbf{70} 043602, (2004).

\bibitem{So}Y.Nishida and  D. T. Son, Phys. Rev. Lett. \textbf{97} 050403, (2006); 

\bibitem{Se}T. Abe and R. Seki,  arXiv:0708.2524, Phys.Rev. \textbf{C79}, 054003, (2009).

\bibitem{Di3}S. Floerchinger et al., arXiv:0808.0150, Phys.Rev. \textbf{B78}, 174528, (2008). 

\bibitem{Go}L. P. Gorkov and T. K. Melik-Barkhudarov, Sov. Phys. JETP \textbf{13}, 1018, (1961).

\bibitem{Lar} A.I. Larkin and Yu. N. Ovchinnikov, JETP \textbf{20} (1965); P. Fulde and R. A. Ferrell, Phys. Rev. A\textbf{135} 550, (1964).

\bibitem{Wil} W.V. Liu and F. Wilczek, Phys. Rev. Lett, \textbf{90} (2003) 047002; E. Gubankova, W. V. Liu and F. Wilczek,
 Phys. Rev. Lett, \textbf{91} 032001, (2003).

G. Sarma, J Phys. Chem.Solid 24, \textbf{24}  1029, (1963).

\bibitem{Bed} P. F. Bedaque et al, Phys. Rev. Lett, \textbf{91} 247002, (2003).

\bibitem{Cal} H. Caldas, Phys. Rev. \textbf{A69} 063602, (2004).

\bibitem{Kr2} B. Krippa, Phys.Lett.\textbf{B643} 104, 2006. 

\end{thebibliography}
\end{document}